\documentclass[prb,superscriptaddress,twocolumn]{revtex4-1}
\usepackage{times,amsmath}
\usepackage{epsfig}
\usepackage{color}
\usepackage{longtable}
\usepackage{graphicx}% Include figure files
\usepackage{dcolumn}% Align table columns on decimal point
\usepackage{bm}% bold math
\usepackage{tabularx}% bold math
\usepackage{hyperref}
\usepackage{multirow}
\usepackage{appendix}
\hypersetup{citecolor=black, filecolor= black, linkcolor= black, urlcolor= black}

\begin{document}
\title{Two-dimensional Obstructed Atomic Insulators with Fractional Corner Charge in MA$_2$Z$_4$ Family}

\author{Lei Wang}
\email{These authors contributed equally to this work}
\affiliation{Shenyang National Laboratory for Materials Science,
Institute of Metal Research, \\Chinese Academy of Science, 110016
Shenyang, Liaoning, People's Republic of China} \affiliation{School
of Materials Science and Engineering, University of Science and
Technology of China,\\Shenyang 110016, People's Republic of China}

\author{Yi Jiang}
\email{These authors contributed equally to this work}
\affiliation{Beijing National Laboratory for Condensed Matter
Physics,  and Institute of Physics, Chinese Academy of Sciences,
Beijing 100190, China}

\author{Jiaxi Liu}
\affiliation{Shenyang National Laboratory for Materials Science,
Institute of Metal Research, \\Chinese Academy of Science, 110016
Shenyang, Liaoning, People's Republic of China} \affiliation{School
of Materials Science and Engineering, University of Science and
Technology of China,\\Shenyang 110016, People's Republic of China}

\author{Shuai Zhang}
\affiliation{Beijing National Laboratory for Condensed Matter
Physics,  and Institute of Physics, Chinese Academy of Sciences,
Beijing 100190, China} \affiliation{University of Chinese Academy of
Sciences, Beijing 100049, China}

\author{Jiangxu Li}
\affiliation{Shenyang National Laboratory for Materials Science,
Institute of Metal Research, \\Chinese Academy of Science,  110016
Shenyang, Liaoning, People's Republic of China} \affiliation{School
of Materials Science and Engineering, University of Science and
Technology of China,\\Shenyang 110016, People's Republic of China}

\author{Peitao Liu}
\affiliation{Shenyang National Laboratory for Materials Science,
Institute of Metal Research, \\Chinese Academy of Science, 110016
Shenyang, Liaoning, People's Republic of China}

\author{Yan Sun}
\affiliation{Shenyang National Laboratory for Materials Science,
Institute of Metal Research, \\Chinese Academy of Science, 110016
Shenyang, Liaoning, People's Republic of China}

\author{Hongming Weng}
%\email{hmweng@iphy.ac.cn}
\affiliation{Beijing National Laboratory for Condensed Matter
Physics, and Institute of Physics, Chinese Academy of Sciences,
Beijing 100190, China} \affiliation{University of Chinese Academy of
Sciences, Beijing 100049, China} \affiliation{Songshan Lake
Materials Laboratory, Dongguan, Guangdong 523808, China}

\author{Xing-Qiu Chen}
\email{xingqiu.chen@imr.ac.cn} \affiliation{Shenyang National
Laboratory for Materials Science, Institute of Metal Research,
\\Chinese Academy of Science, 110016 Shenyang, Liaoning, People's
Republic of China} \affiliation{School of Materials Science and
Engineering, University of Science and Technology of
China,\\Shenyang 110016, People's Republic of China}

\date{\today}
\begin{abstract}
According to topological quantum chemistry, a class of electronic materials
have been called obstructed atomic insulators (OAIs), in which a portion of
valence electrons necessarily have their centers located on some empty
\emph{Wyckoff} positions without atoms occupation in the lattice.
The obstruction of centering these electrons coinciding with their host atoms
is nontrivial and results in metallic boundary states when the boundary is properly cut.
Here, on basis of first-principles
calculations in combination with topological quantum chemistry analysis,
we propose two dimensional MA$_2$Z$_4$ (M = Cr, Mo and W; A = Si and Ge, Z = N, P and As) monolayer family are all OAIs.
A typical case is the recently synthesized MoSi$_2$N$_4$.
Although it is a topological trivial insulator with the occupied electronic
states being integer combination of elementary band representations, it has valence
electrons centering empty \emph{Wyckoff} positions. It exhibits unique OAI-induced
metallic edge states along the (1\={1}0) edge of MoSi$_2$N$_4$ monolayer and
the in-gap corner states at three vertices of certain hexagonal nanodisk samples respecting C$_3$ rotation symmetry.
The readily synthesized MoSi$_2$N$_4$ is quite stable and has a large bulk band gap of 1.94 eV,
which makes the identification of these edge and corner states most possible for experimental clarification.
\end{abstract}

\maketitle

\emph{Introduction}. --Topological materials including topological
insulators and topological semimetals have attracted intensive
attentions~\cite{review-RMP-TI,review-RMP-TIS,QSHE-HgTe-science,hsieh_tunable_2009,
zhang_topological_2009,Hsien-Nature-2008,HgTe-2007,
xiang_quantum_2016,Rb3O-topology,TiB4-PRB,QSHE-silicene-PRL,
Na3Bi-PRB,Co3Sn2S2-Science,Be-NLSs-PRL,Cd3As2-Dirac-2013,
TaAs-PRX,SnTe-TCI-NC-2012,TBG-topo-PRL,HT-Weyl-science-bulletin,TCI-Fu-PRL,
Schindler-SA-2018,Ezawa-PRB-2018,Schindler-NP-2018,Yue-NP-2019,Xu-PRL-2019,
Xu-PRL-2019,Wang-PRL-2019,Sheng-PRL-2019,NanoL-2019,NanoL-2022,ChenC-PRL-2020,
PRB-yao-2021}, mainly due to their
nontrivial bulk band dispersion and metallic surface (or edge)
states. In recent years, the development of topological quantum
chemistry (TQC)~\cite{TQC-2017,MTQC-2021,TMM-2021} and symmetry
indicator~\cite{SI-2017,SIM-2018} provides convenient and efficient tools to
the high-throughput discoveries of topological quantum
materials~\cite{Class1-Nature-2019,Class2-Nature-2019,
Class3-Nature-2019,ClassM-Nature-2020,ClassM-arxiv-2021}. Within the
TQC theory, topological trivial insulators can be defined by band
representations (BRs) of valence bands, which are equivalent to a
set of exponentially localized Wannier
functions~\cite{EBR-2018-PRB}. Any BRs can be given by a linear
combination of elementary band representations (EBRs), which are induced from irreducible representations at maximal \emph{Wyckoff}
positions (WPs)~\cite{EBR-2018-PRB,Double-2017-JAC,Graph-2017-PRE}. If
some of the coefficients of linear combination of elementary band
representations (LCEBRs) are rational fractions, the material
is a stable topological insulator or topological semimetal. If all the LCEBRs of a material exhibit non-negative integer coefficients, the materials is then a topological trivial insulator~\cite{OAI-2021-6}.

Recently, a series of unique topologically trivial insulators,
dubbed obstructed atomic insulators (OAIs), have been found~\cite{OAI-2021-6,OAI-2021-11}. In difference from the BRs of
atomic insulators induced from the atomic orbitals locating at
atom-occupied \emph{Wyckoff} positions (AOWPs), the BRs of OAIs are
induced from additional atom-unoccupied \emph{Wyckoff} positions (AUWPs). In other words, for atomic insulators electrons fill atomic orbitals at the AOWPs, but for OAIs a portion of electrons have to occupy those at the AUWPs. Note that in any periodic lattice of solid materials there are many so-called AUWPs, which can be easily
identified via the international tables for crystallography. Nevertheless, not all AUWPs are necessary for OAIs, while only
those electron-filled AUWPs are necessary for OAIs. Such electron-filled
AUWPs for OAIs are also named as obstructed Wannier charge centers
(OWCCs)~\cite{OAI-2021-6}. Importantly, when the cleavage termination cuts
through those electron-filled AUWPs (namely, OWCCs) in an OAI, the
metallic surface states will emerge. This crucial feature makes OAIs
be potential candidates for superconductivity and
catalysis\cite{OAI-2021-11,unconven-M-2021,app-electride-2021-PRB,coupling-Corner-states-2020,FFJD-2022}.
To date, OAIs including 3,383 paramagnetic and 30 magnetic materials are reported in three-dimensional (3D) materials
\cite{OAI-2021-6,OAI-2021-11}.
However, 2D OAIs, which could be used in low dimensional devices\cite{HOTI-OAI,2D-SWI-2022}, have not been systematically investigated.

%Furthermore, although 2D OAIs with inversion symmetry or PT symmetry are founded in graphdiyne \cite{HOTI-OAI} and liganded Xenes

%\cite{2D-SWI-2022}, respectively, whether 2D OAIs can be found in the system without inversion symmetry remains unexplored.
%although high-order topological insulators or
%and semimetals have been discussed in-depth \cite{Schindler-SA-2018,Ezawa-PRB-2018,
%Schindler-NP-2018, Yue-NP-2019, Xu-PRL-2019,Wang-PRL-2019, Zhang-PRL-2020,
%Sheng-PRL-2019,HOTI-OAI,NanoL-2019,NanoL-2022,ChenC-PRL-2020,HOTCI-MoS2,PRB-yao-2021},

\begin{figure*}
\begin{center}
\includegraphics[width=0.8\textwidth]{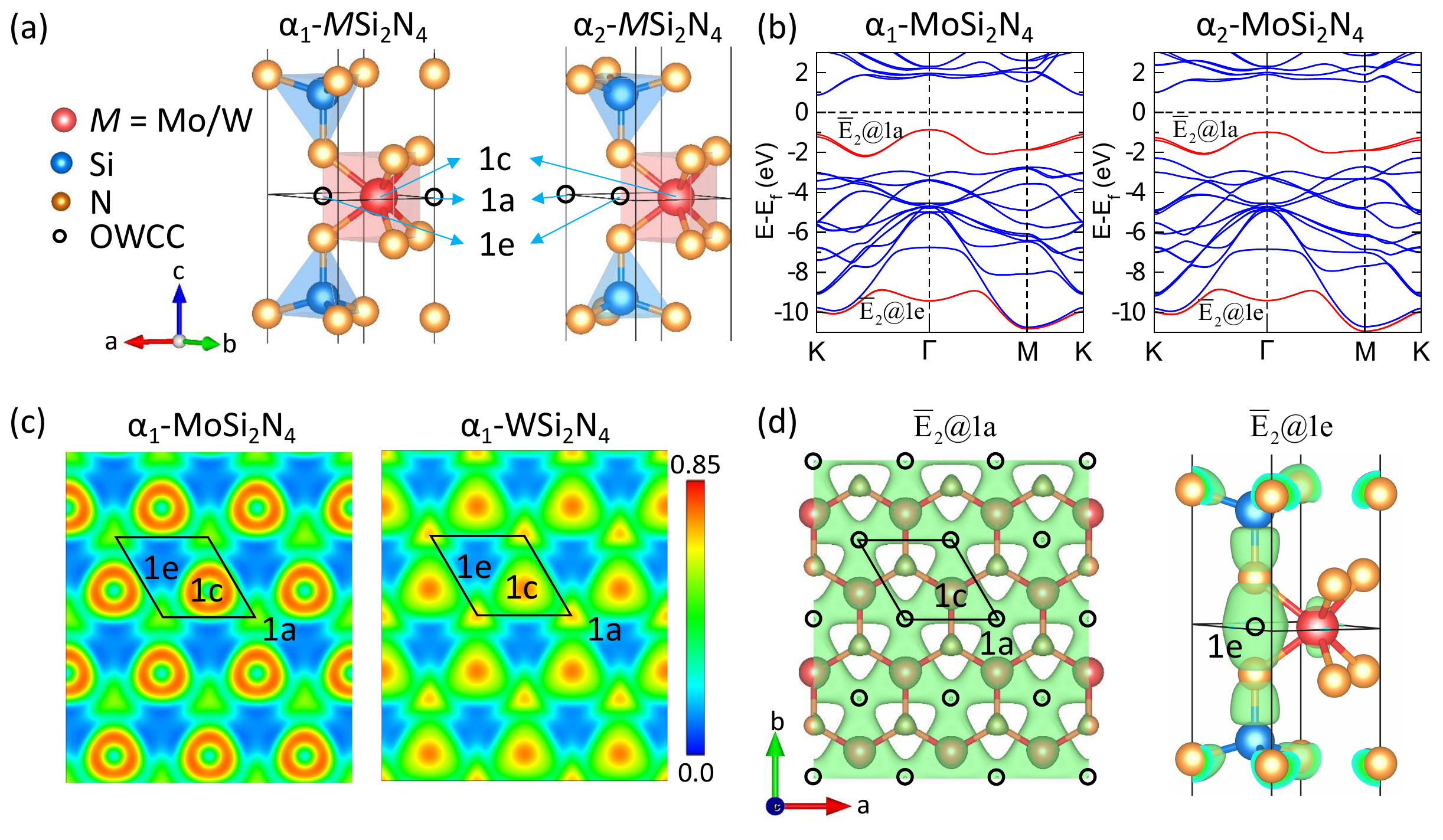}
\caption{(Color online) Lattice and electronic structures of
\emph{M}Si$_2$N$_4$ monolayer (\emph{M} = Mo, W). (a) Lattice
structures of $\alpha_1$- and $\alpha_2$-\emph{M}Si$_2$N$_4$
monolayer, where the black circle denotes the obstructed wannier
charge center (OWCC). (b) The electronic band structures of
$\alpha_1$- and $\alpha_2$-MoSi$_2$N$_4$ monolayer with the inclusion of spin-orbit coupling, where the red
bands correspond to the band representation of \={E}$_2$@1a and
\={E}$_2$@1e. (c) The electron localization functions (ELF) of
$\alpha_1$-MoSi$_2$N$_4$ and $\alpha_1$-WSi$_2$N$_4$ monolayer.
1$a$, 1$c$, and 1$e$ are \emph{Wyckoff} positions, and the localized
charges occupy at the 1$a$ and 1$c$ \emph{Wyckoff} positions. The
solid lines denote the primitive cell of
$\alpha_1$-\emph{M}Si$_2$N$_4$ monolayer materials. (d) The charges
distribution of the band representation \={E}$_2$@1a and
\={E}$_2$@1e of $\alpha_1$-MoSi$_2$N$_4$. The green hook face is the
isosurface with value of 0.008 e/\AA$^3$ and the black circles denote the
OWCCs at both 1$a$ and 1$e$.  In the left panel of (d), to show 1$a$ position clearly, the top and bottom Si-N layers of $\alpha_1$-MoSi$_2$N$_4$ monolayer are removed.} \label{fig1}
\end{center}
\end{figure*}

Recently, 2D monolayer MA$_2$Z$_4$ family~\cite{Hong-Science-2020,
Wang-NC-2021} with septuple-atomic-layer lattices have been
experimentally or theoretically reported. This MA$_2$Z$_4$ family
crystallizes in five different crystalline phases, including 72
theoretically suggested stable materials~\cite{Wang-NC-2021}. By
inspecting the electronic structure of theoretically predicted stable monolayer $\alpha_1$-WSn$_2$N$_4$ semiconductor with an indirect band gap of 0.18 eV, we found that it possesses the typical inverted energy band between W-$d$$_{z^2}$ and N-$p$$_z$ orbitals at the centered
$\Gamma$ point of the Brillouin Zone (BZ) around the band gap (Appendix Fig. A1). Although the band inversion typically signals to have nontrivial topological nature, both its
topological indictor and mirror chern number equal to zero, indicating a topological trivial insulator. Furthermore, by deriving its electronic band structure of the edge boundary, metallic edge states nevertheless occur, which is very similar to the metallic surface states of 3D OAIs ~\cite{OAI-2021-6,OAI-2021-11}. Interestingly, we further revealed that in total 16 monolayer MA$_2$Z$_4$
semiconductors with 34 valence electrons (VEC) including
experimentally synthesized $\alpha_1$-MoSi$_2$N$_4$ exhibit similar
electronic band features to $\alpha_1$-WSn$_2$N$_4$ and they all are topological trivial insulators. Their special
electronic structures make us think over whether these monolayer
MA$_2$Z$_4$ semiconductors are 2D OAIs. If yes, any novel properties will emerge in
these 2D OAI family?

With this motivation, by means of first-principles calculations in
combination with TQC analysis, we report 16 monolayer 34-VEC
MA$_2$Z$_4$ semiconductors as 2D OAIs. By taking the
experimentally synthesized MoSi$_2$N$_4$ semiconductor as a typical
example, we identify the occurrence of the localized charge at the AUWPs as the crucial fingerprint of OAIs. Additionally, the metallic edge states appear in the band gap along the (1\={1}0) direction on the boundary of MoSi$_2$N$_4$ monolayer at which cleavage terminations exactly cut through OWCCs. Interestingly, in-gap corner states found in second-order topological insulators also occur in the C$_3$-symmetric hexagonal nanodisk of MoSi$_2$N$_4$.

\begin{table*}[htp]
\setlength{\tabcolsep}{3.8mm}
\begin{center}
\caption{All possible decompositions of the BRs of
$\alpha_1$-MoSi$_2$N$_4$ and $\alpha_2$-MoSi$_2$N$_4$ into linear
combinations of the EBRs in the double space group P$\bar{6}$m2 (No.
187). The first row gives the possible EBRs induced from different
orbitals at the \emph{Wyckoff} positions 1$a$, 1$c$, and 1$e$; the
numbers below are the multiplicities of each EBR in the
corresponding decomposition. Note that only a portion of LCEBRs are
listed here and the complete LCEBRs are given in the Appendix Table A1
and Table A2.} \label{lCEBR}
\begin{tabular}{cccccccccc}
\hline\hline
\rule{0pt}{10pt}
Compounds &  \={E}$_1$@1a & \={E}$_2$@1a & \={E}$_3$@1a  & \={E}$_1$@1c & \={E}$_2$@1c & \={E}$_3$@1c  & \={E}$_1$@1e & \={E}$_2$@1e & \={E}$_3$@1e   \\\hline
\multirow{6}{*}{$\alpha_1$-MoSi$_2$N$_4$}
  &6 &7 &5 &1 &0 &0 &1 &1 &0 \\
  &5 &6 &4 &2 &1 &1 &1 &1 &0 \\
  &4 &5 &3 &3 &2 &2 &1 &1 &0 \\
  &3 &4 &2 &4 &3 &3 &1 &1 &0 \\
  &2 &3 &1 &5 &4 &4 &1 &1 &0 \\
  &1 &2 &0 &6 &5 &5 &1 &1 &0 \\ \hline

\multirow{6}{*}{$\alpha_2$-MoSi$_2$N$_4$}
 &5 &6 &5 &2 &1 &0 &1 &1 &0  \\
 &4 &5 &4 &3 &2 &1 &1 &1 &0  \\
 &3 &4 &3 &4 &3 &2 &1 &1 &0  \\
 &2 &3 &2 &5 &4 &3 &1 &1 &0  \\
 &1 &2 &1 &6 &5 &4 &1 &1 &0  \\
 &0 &1 &0 &7 &6 &5 &1 &1 &0  \\
\hline\hline
\end{tabular}
\end{center}
\end{table*}

\emph{Lattice structure of MoSi$_2$N$_4$ monolayer}. --- The
experimentally synthesized $\alpha_1$-MoSi$_2$N$_4$
~\cite{Hong-Science-2020} with 34 VEC belongs to MA$_2$Z$_4$
family~\cite{Wang-NC-2021}. In our theoretical
predictions~\cite{Wang-NC-2021}, there are 15 monolayer 34-VEC
MA$_2$Z$_4$ semiconductors, which mainly crystallize in two classes
of lattice structures, $\alpha_1$- and $\alpha_2$-MA$_2$Z$_4$. For
MoSi$_2$N$_4$, the formation energy of $\alpha_1$-MoSi$_2$N$_4$ is
24 meV/atom lower than that of $\alpha_2$-MoSi$_2$N$_4$. Similarly,
$\alpha_1$ phase of WSn$_2$N$_4$ is energetically more stable
by 1.3 meV/atom than its $\alpha_2$ phase. Both $\alpha_1$- and
$\alpha_2$-MoSi$_2$N$_4$ have the hexagonal lattices with the atomic
sequence of N-Si-N-Mo-N-Si-N (the space group of P$\bar{6}$m2 (No.
187)\footnote{For a MoSi$_2$N$_4$ monolayer, it should be described
by layer group of p$\bar6$m2 (No. 78). In order to facilitate
the reference of \emph{Wyckoff} positions from the international tables
for crystallography and elementary band representations from https://www.cryst.ehu.es/,
the space group of P$\bar{6}$m2 (No. 187) is adopted here, which is also rational.}), as shown in Fig. ~\ref{fig1}(a). This septuple-atom-layer can
be viewed by inserting 2H-MoS$_2$-type MoN$_2$ into
$\alpha$-InSe-type SiN. In $\alpha_1$-MoSi$_2$N$_4$ there are two
different nitrogen WPs of 2$g$ (0, 0, $\pm z$) and 2$i$ (2/3, 1/3,
$\pm z$), whereas for $\alpha_2$-MoSi$_2$N$_4$ two different
nitrogen WPs of 2$h$ (1/3, 2/3, $\pm z$) and 2$i$ (2/3, 1/3, $\pm
z$). Thus, here we denote three different nitrogen WPs as N$_g$,
N$_i$, and N$_h$ for sake of convenient discussion below. As
illustrated in Fig.~\ref{fig1}(a), both N$_g$ for
$\alpha_1$-MoSi$_2$N$_4$ and N$_h$ for $\alpha_2$-MoSi$_2$N$_4$
locate at the top or the bottom of the monolayer MoSi$_2$N$_4$, but
N$_i$ atoms sit in the layer between Si of 2\emph{i} (2/3,
1/3, $\pm z$) and Mo of 1\emph{c} (1/3, 2/3, 0). Thus,
nitrogen atoms form two basic local structures, a tetrahedron
consisting of three N$_g$ (or N$_h$) atoms and one N$_i$ atom, and a
triangular prism composed of six N$_i$ atoms. Si and Mo atoms sit at
the centering positions of the tetrahedron and the prism,
respectively. It needs to be emphasized that N$_i$ and Si atoms have
the same \emph{Wyckoff} notations of 2$i$ (2/3, 1/3, $\pm z$) but
with different $z$ values.

\emph{OAI identification of MoSi$_2$N$_4$ monolayer}. ---
$\alpha_1$-MoSi$_2$N$_4$ is an indirect semiconductor with an
experimental band gap of 1.94 eV, comparable to the DFT-derived gap
of 1.74 eV (PBE) and 2.30 eV (HSE06)
\cite{Wang-NC-2021}. It is also a topologically trivial insulator,
according to topological analysis using mirror chern number. Conceptually, the occupied electronic
bands of 3D topological trivial insulator can be expressed by a
non-negative integer linear combination of EBRs using the
TQC theory~\cite{TQC-2017,MTQC-2021,TMM-2021}. Since
MoSi$_2$N$_4$ monolayer is a 2D material, only the band
representations (BRs) of high symmetry points $\Gamma$, M and K at
k$_z$=0 plane are considered. And since the EBRs of \emph{Wyckoff} positions 2$g$, 2$h$ and 2$i$ can be obtained by the combinations of
EBRs of \emph{Wyckoff} positions 1$a$, 1$c$, and 1$e$, the maximum
\emph{Wyckoff} positions 1$a$, 1$c$, and 1$e$ are chosen to perform the
EBRs decomposition. Therefore, the LCEBRs of the occupied electronic
bands of $\alpha_1$- and $\alpha_2$-MoSi$_2$N$_4$ have been
derived in Table~\ref{lCEBR}. Note that we only listed a
portion of LCEBRs, and the complete LCEBRs of $\alpha_1$- and
$\alpha_2$-MoSi$_2$N$_4$  can be found in Appendix Table A1 and
Table A2, respectively. Interestingly, for $\alpha_1$-MoSi$_2$N$_4$,
the results of all the LCEBRs show that there are six EBRs
(\=E$_1$@1$a$, \=E$_2$@1$a$, \=E$_1$@1$c$, \=E$_2$@1$c$,
\=E$_1$@1$e$ and \=E$_2$@1$e$) with the non-zero integer
combination. This fact means these six EBRs can not be decomposed
and they have to be linked to electron-filled \emph{Wyckoff} positions of 1\emph{a},
1\emph{c}, and 1\emph{e}. The 1\emph{c} site is occupied by Mo atom
whereas the 1\emph{a} and 1\emph{e} sites are \emph{null} without
any atomic occupation. In terms of the OAI
definition~\cite{OAI-2021-6,OAI-2021-11}, $\alpha_1$-MoSi$_2$N$_4$
is an OAI and the 1\emph{a} and 1\emph{e} sites are the OWCC.
Similarly, the non-zero integer of LCEBR of the 1\emph{a} and
1\emph{e} AUWPs indicates that $\alpha_2$-MoSi$_2$N$_4$ is also an
OAI, and the 1\emph{a} and 1\emph{e} AUWPs are the OWCC.

For the OAI feature, the most key point is to check whether the
AUWPs have the localized charges (namely, electron filling). To
elucidate the real-space charge localizations of MoSi$_2$N$_4$ we
have thus visualized the electron localization function (ELF)
(Fig.~\ref{fig1}(c)) on the centering Mo-atom layer in which four
indecomposable EBRs correspond to both 1$a$ (\emph{null}) and
1$c$ (Mo) AUWP. It can be seen that the charges obviously localize
at the 1$c$ Mo site and the $null$ 1$a$ AUWP. The feature is more
apparent in $\alpha_1$-WSi$_2$N$_4$ and $\alpha_2$-MoSi$_2$N$_4$
(see Appendix Fig. A2). We find that the
localized charges at the 1\emph{c} AOWP are mainly contributed by Mo atoms,
whereas the ones at the \emph{null} 1\emph{a} AUWP originate
from the orbital hybridizations between Mo and N$_i$ atoms, which is
in good agreement with obstructed atomic limit for SSH model with
$sp$ orbital hybridization\cite{TQC-2017}. Furthermore, in Fig.~\ref{fig1}(d)
we plot the partial charge densities of the two indecomposable
EBRs responsible for the isolated electronic bands marked by
\=E$_2$@1$a$ and \=E$_2$@1$e$ in Fig.~\ref{fig1}(b), evidencing
the distribution of charges at both the 1$a$ and 1$e$ AUWPs.

Following the similar analysis, we have checked another 15
MA$_2$Z$_4$ monolayer semiconductors with 34 valence electrons predicted in our
previous work~\cite{Wang-NC-2021}. The results demonstrate that all
of them are typical OAIs, as summarized in Appendix Table A3.

\emph{Obstructed metallic edge states and in-gap corner states}. ---
Physically, topological nontrivial materials with $d$-1 dimensional
boundary states are defined as first-order topological materials (where $d$ is the dimension of topological materials). The ones with $d$-$n$ ($n$ $>$ 1) dimensional boundary states are the $n$th order topological materials. Interestingly, In similarity with 3D
topological insulators, for a 3D OAI the $d$-1 dimensional metallic
surface states occur on the 2D surface with the cleavage
terminations exactly cutting through an
OWCC~\cite{OAI-2021-6,OAI-2021-11}.
%And For 2D OAIs, both the $d$-1 dimensional metallic
%surface states and $d$-2 dimensional in-gap corner states are reported.
Hence, it is desirable to examine to see whether or not the metallic edge states of the 1D boundary occur for the 2D OAI of monolayer MoSi$_2$N$_4$.

\begin{figure}
\begin{center}
\includegraphics[width=0.48\textwidth]{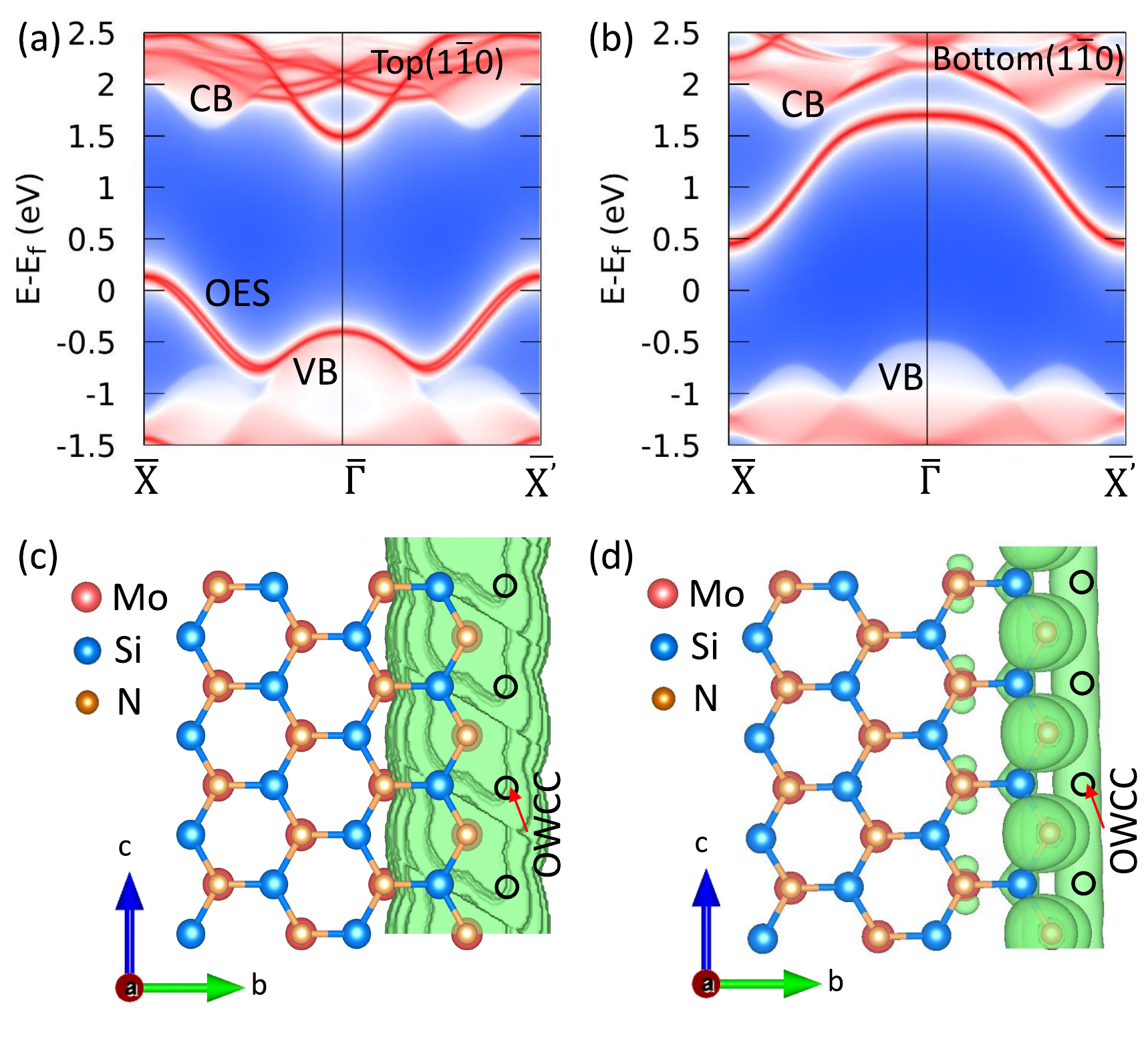}
\caption{(Color online) Obstructed metallic edge states of
$\alpha2$-MoSi$_2$N$_4$ monolayer. (a-b) The projected edge states along the
(1\={1}0) direction of the boundary with cutting through the OWCC at 1$a$ site. CB
and VB are the conduction bands and valence bands. OES is the
obstructed metallic edge states. (c) Charge distributions of metallic OES at Fermi level shown in (a) derived by Wannier-based Hamiltonian of $\alpha_2$-MoSi$_2$N$_4$ nanoribbon. (d) Charge distributions of metallic OES at Fermi level shown in (a) derived by first-principles calculations of $\alpha_2$-MoSi$_2$N$_4$ nanoribbon. The black circle marked by red
arrow are OWCC. Green hook face is the isosurface of the charge
of OES with a value of 0.002 e/\AA$^3$. } \label{fig2}
\end{center}
\end{figure}

\begin{figure*}
\begin{center}
\includegraphics[width=0.85\textwidth]{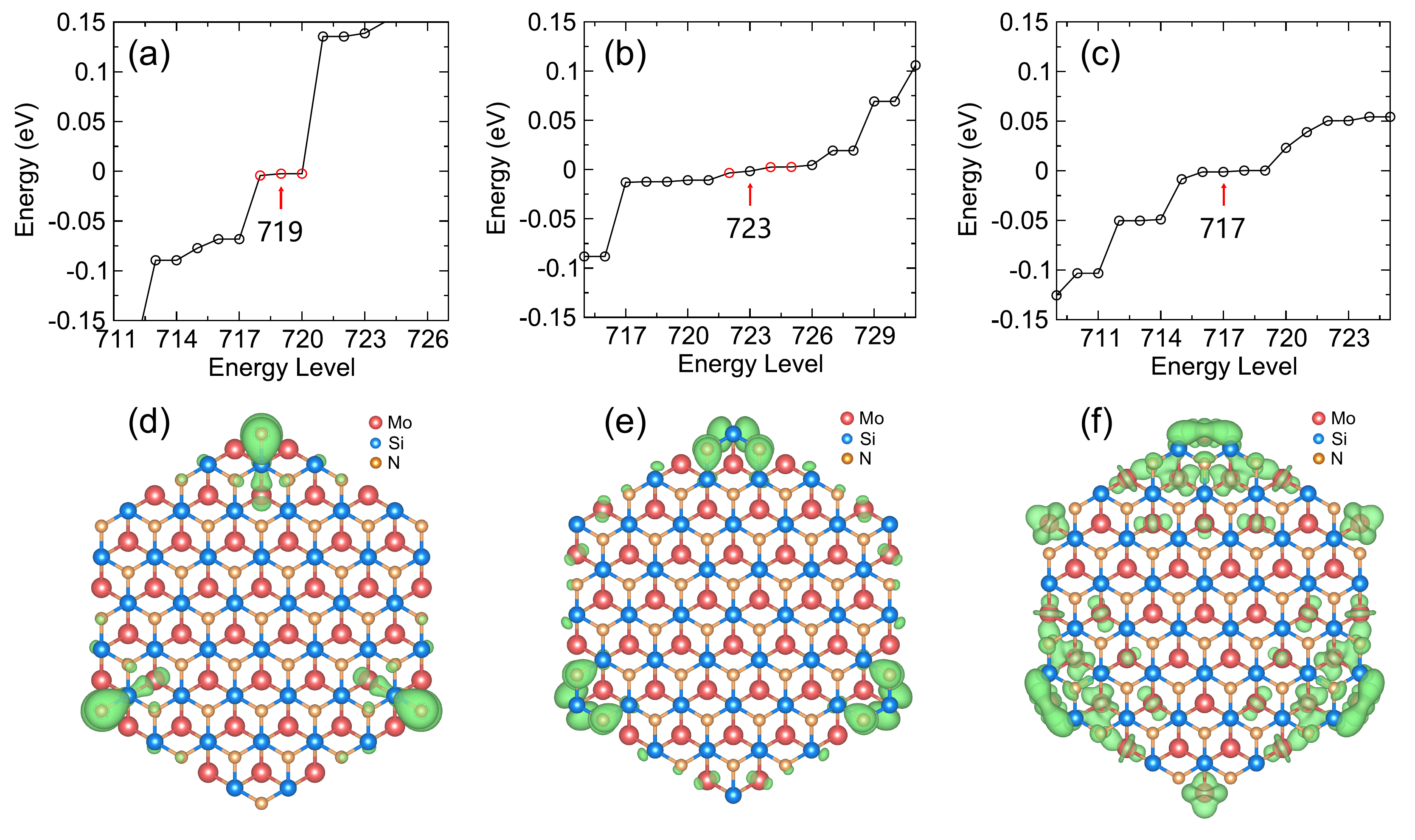}
\caption{(Color online) (a-c) Energy spectra of C$_3$-symmetric
hexagonal-shaped nanodisk of $\alpha_1$-MoSi$_2$N$_4$ with the N, Si, and Mo atom in the center, where the
occupied energy level is marked by the red arrow and red circles
represent corner states. (d-f) The charge distributions of the 718,
722, and 716 energy levels appearing in (a)-(c), respectively.}
\label{fig3}
\end{center}
\end{figure*}

Note that because for $\alpha_1$-MoSi$_2$N$_4$ its two N$_g$ and two
N$_i$ atoms are exactly above its OWCC-type 1$a$ (0, 0, 0) and 1$e$
(2/3, 1/3, 0) AUWPs, it is hard to cut an edge that only contains
the OWCC but without containing the N atoms.
However, this difficulty does not occur in its $\alpha_2$ phase of
OAI, since no atoms are exactly above its OWCC-type 1$a$ AUWP (see
Fig.~\ref{fig1}(a) and Table~\ref{lCEBR}). On basis of Wannier-based
Hamiltonian of $\alpha_2$-MoSi$_2$N$_4$, we derived the
electronic band structures of the edge boundary along (1\={1}0)
direction, which only cuts through the OWCC-type 1$a$ (0, 0, 0) AUWP
(Fig. \ref{fig2}(a-c)). Notably, the apparent obstructed metallic
edge states (OESs) occur between the conduction bands (CB) and
valence bands (VB) of $\alpha_2$-MoSi$_2$N$_4$ monolayer.
%By counting the number of the bands from the lowest occupied band to a
%band to guarantee the charge neutrality of the system, we marked
%the valence and conduction bands by the blue and red curves in Fig.
%\ref{fig2}(c). According to Fig. \ref{fig2}(c), it can be seen that
%the OESs in Fig. \ref{fig2}(a) are half-filled and the edge boundary
%is metallic.
This phenomenon agrees well with the observations in 3D OAIs
\cite{OAI-2021-6,OAI-2021-11}. By calculating the charge
distribution of the OESs at the Fermi level within a nanowire model
with a width of 20 unit cells, the charges localize around the
OWCC-type 1$a$ AUWP and the decay of the OESs only has a depth of
about 0.4 nm, as seen in Fig. \ref{fig2}(c). Such result is also verified by
the first-principles calculations, in Fig. \ref{fig2}(d).

%In 2D materials, zero-energy and zero-dimensional in-gap states are
%reported in their second-order topological insulator (SOTI) phase.
%For instance, the monolayer graphdiyne
%~\cite{Sheng-PRL-2019,HOTI-OAI}, $\gamma$-Graphyne
%~\cite{NanoL-2019,NanoL-2022}, Bi/EuO(111) ~\cite{ChenC-PRL-2020},
%atomically thin transition metal dichalcogenides ~\cite{HOTCI-MoS2},
%and other haxagonal lattice materials\cite{PRB-yao-2021} are
%recently identified as SOTI, and there are six zero-dimensional
%in-gap corner states in their hexagonal-shaped nanodisk, related to
%a C$_6$ symmetry. Such in-gap corner states can be ascribed to their
%$filling$ $anomaly$ \cite{HOTI-OAI}, which is indeed consistent with
%that of OAI. The key feature for SOTI is that even though the in-gap
%states are merged into the bulk states, the nontrivial band topology
%manifests in the $filling$ $anomaly$ \cite{HOTI-OAI}. Thus,
%two-dimensional SOTI with in-gap corner states is also OAI
%\cite{HOTI-OAI}.

Next we turn to checking whether or not the 2D OAI of monolayer
MoSi$_2$N$_4$ has the $d$-2 dimensional (namely zero-dimensional (0D)) in-gap corner states. We derived the
electronic structures of the 0D nanodisk modeling of
$\alpha_1$-MoSi$_2$N$_4$. By analyzing its structural details (see
Appendix Fig. A3, Fig. A4, and Table A4), we constructed a
C$_3$-symmetric triangle and hexagonal nanodisk
with the armchair edge to keep the electrically neutral
stoichiometric ratio of 1:2:4 over Mo:Si:N. Since monolayer
MoSi$_2$N$_4$ holds three types of maximum \emph{Wyckoff} positions,
there are three possible geometries for both triangle and hexagonal
nanodisks by varying the type of atom at center site. No matter which
type of atom sits at the center of the C$_3$-symmetric triangle or
hexagonal nanodisk, their stoichiometric ratio is always remained by
removing the center atoms (see Appendix Table A4).

%In difference from the above monolayer graphdiyne
%~\cite{Sheng-PRL-2019,HOTI-OAI}, $\gamma$-Graphyne
%~\cite{NanoL-2019}, and Bi/EuO(111) ~\cite{ChenC-PRL-2020} which are
%topologically nontrivial, monolayer MoSi$_2$N$_4$ is an OAI with a
%trivial topology.

To argue whether the monolayer MoSi$_2$N$_4$ holds the in-gap
corner states, one has to see the so-called gapped edge states on the
boundaries and the in-gap corner states siting the vertexes of the
C$_3$-symmetric triangle or hexagonal nanodisk.
However, depending on the $filling$ $anomaly$ at the center site, the in-gap corner states certainly occur in C$_3$-symmetric triangle and hexagonal nanodisk.

Recently, a general formula to calculate the corner charges of the C$_n$ symmetry nanodisk was developed
\cite{corner-Cn-2021PRB,corner-Cn-2019PRB,corner-PRR}, as follows,
\begin{equation}\label{Qc}
Q_{\text {corner}}=Q_{c_{-} X}^{(n)} \equiv
\frac{\left(n_{X}^{(\text {ion})}-n_{X}^{(\mathrm{e})}\right)|e|}{n}
\quad(\bmod |e|),
\end{equation}
where $Q_{c_{-} X}^{(n)}$ means the corner charge when the center of
the crystal locates at the $X$ site, $n$ is the fold of rotation
axis. $n_{X}^{(\text {ion})} |e|$ and $n_{X}^{(\mathrm{e})}$ are the
ionic charges and the number of electronic Wannier functions at the
$X$ site, respectively. Applying Eq.~\ref{Qc} to monolayer
MoSi$_2$N$_4$, we obtained the corner charge of the
C$_3$-symmetric triangle or hexagonal nanodisk of $Q_{1a}^{(n)}$ =
1/3 $|$e$|$, $Q_{1e}^{(n)}$ = 2/3 $|$e$|$ and $Q_{1c}^{(n)}$ = 0
$|$e$|$. Here, the 1$a$ , 1$e$ and 1$c$ sites are responsible for
the N, Si and Mo atoms at the center of the $C_3$-symmetric
crystals. Furthermore, we derived the energy spectra and charge distributions of our
constructed triangle or hexagonal nanodisk. Note that the energy
spectra and charge distribution of the N-centered hexagonal
nanodisk are further compiled in Fig.~\ref{fig3}(a, d), while the
others are given in Appendix Fig. A5. It can be seen that three
corner states at the N atoms meeting the $C_3$ rotational symmetry
are in-gap and nearly zero-energy, labeling Nos. 718, 719 and 720. The other three
corners without any charge distribution can be actually viewed as
charge neutral "edges". The states of Nos. 718 and 719 are occupied
and the No. 720 states are unoccupied. Because this system is spinless and the number of
electrons of this system is 1438, we need to add two or subtract
four electrons to make this system fully
occupied or fully empty so that the system is gapped. The added two or
subtracted four electrons correspond to -2/3 $|$e$|$ and 4/3 $|$e$|$
corner charge at its each $C_3$-symmetry corner. This result is
equivalent to the calculated 1/3 $|$e$|$ through Eq.~\ref{Qc}.
We note that the same conditions exist in the corner charge
calculations of both Si-centered and Mo-centered hexagonal
nanodisks. In Si-centered hexagonal nanodisk, three corner states
are separated by one edge state (see Figs.~\ref{fig3}(b, e) and
Appendix Fig.~A6). The corner state below edge state can be pushed into the gap of edge
via an appropriate edge potential\cite{TQC-2017}. In Mo-centered hexagonal nanodisk, the
states near the Fermi level are distributed not only at three
corners in similarity to those of N-centered hexagonal nanodisk, but
also at three charge neutral "edges". Therefore, no corner states
occur in their energy level near the Fermi level, which reflects well
the zero corner charge of the Mo-centered hexagonal nanodisk (see
Figs.~\ref{fig3}(c, f) and Appendix Fig. A7).
%Therefore, for MoSi$_2$N$_4$ monolayer of OAI, we call it second-order OAI, due to the existence of in-gap corner states, namely $d$-2 dimensional in-gap states.

\emph{Conclusions}. ---
To summarize, we have identified the 16 MA$_2$Z$_4$ monolayer family materials with 34 valence electrons as 2D obstructed atomic insulators. They are featured by the occurrence of half-filled obstructed metallic edge states in 1D  nanowires and in-gap corner states in 0D C$_3$-symmetric hexagonal nanodisks. Moreover, the 2H-MoS$_2$ monolayer and $\alpha$-InSe monolayer, the two basic constituent units for the MA$_2$Z$_4$ monolayer family, are also identified as obstructed atomic insulators.
Our work proposes a promising realization of 2D obstructed atomic insulators without inversion symmetry and provides a new platform to explore exotic phases of condensed matter and their associated novel properties.

%\begin{acknowledgments}
\emph{Acknowledgments}. ---  Work was supported by the National
Science Fund for Distinguished Young Scholars (grant number
51725103), the National Natural Science Foundation (Grant No. 11925408, 11921004 and 12188101), the Ministry of Science and Technology of China (Grant No. 2018YFA0305700), the Chinese Academy of Sciences (Grant No. XDB33000000), the K. C. Wong Education Foundation (GJTD-2018-01), and the Informatization Plan of Chinese Academy of Sciences(Grant No. CAS-WX2021SF-0102). All calculations have been performed on the
high-performance computational cluster in the Shenyang National
University Science and Technology Park.

%\end{acknowledgments}
%\bibliography{ref}

\begin{appendix}
\renewcommand\thefigure{\Alph{section}\arabic{figure}}
\renewcommand\thetable{\Alph{section}\arabic{table}}

\section{Computational methods}
Vienna ab $initio$ simulation package (VASP)~\cite{kresse-PRB-1996,kresse-PRB-1999}
with exchange-correlation potential of Perdew-Burke-Ernzerhof (PBE) and projector augmented wave (PAW) method was used to perform first-principles calculations. 20 {\AA} vacuum was set to exclude the interactions between the layers with periodic images. 500 eV cutoff energy and 15 $\times$ 15 $\times$ 1 $k$-mesh in $\Gamma$-centered Monkhorst-Pack scheme were chosen in self-consistent calculation process. The structure was optimized until the force and energy less than 10$^{-3}$ eV/{\AA} and 10$^{-6}$ eV/{\AA}, respectively. All the C$_3$-symmetric nanodisks were calculated by only $\Gamma$ points. In addition, combating with Wannier90~\cite{wannier90} code, density functional theory (DFT)-derived Wannier functions (WFs) was constructed. And the Hamiltonians for each compounds were derived. Iterative Green functions method \cite{GreenF} was applied to calculate the semi-infinite spectral function.

\setcounter{figure}{0}
\setcounter{table}{0}

\begin{figure*}[htp]
\begin{center}
\includegraphics[height=0.70\textwidth]{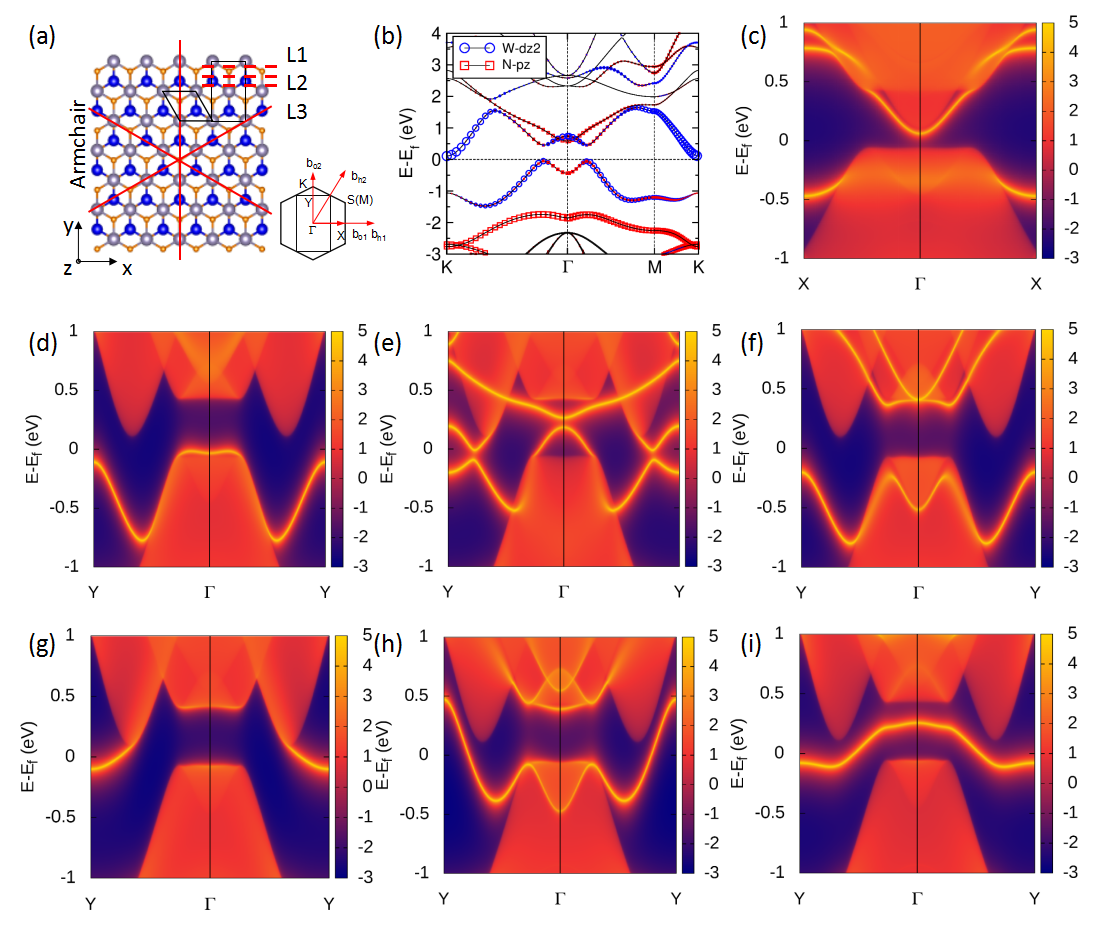}
\caption{(a) The top view of constructed orthorhombic cell of $\alpha_2$-WSn$_2$N$_4$. The armchair edge is paralleled to
mirror $M_x$. L$i_{up}$ and L$i_{down}$ ($i$= 1, 2, 3) are the edge above or below
line L$_i$. The insert shows the Brillouin zone of orthorhombic
and hexagonal cell. (b) The band structure of $\alpha_2$-WSn$_2$N$_4$. The blue circles and
red square are the weight of W$_{dz^2}$ and N$_{p_z}$.
The spectral function for (c) armchair, (d) L1$_{up}$, (e) L1$_{down}$, (f) L2$_{up}$, (g) L2$_{down}$,
(h) L3$_{up}$ and (i) L3$_{down}$ edges. The bright gold, red and navy
part represent surface states, bulk states and vacuum, respectively.}
\label{alpha1-WSn2N4}
\end{center}
\end{figure*}

\begin{figure*}[htp]
\begin{center}
\includegraphics[height=0.50\textwidth]{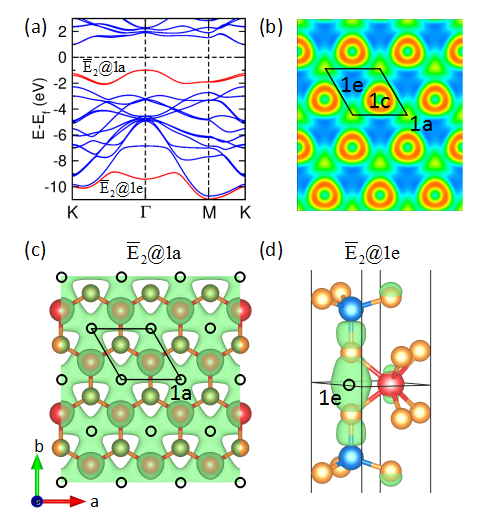}
\caption{ Electronic structure of $\alpha_2$-MoSi$_2$N$_4$ monolayer. (a) The band structure of $\alpha_2$-MoSi$_2$N$_4$ monolayer, where the red bands correspond to the band representation of \={E}$_2$@1a and \={E}$_2$@1e. (b) The ELF of $\alpha_2$-MoSi$_2$N$_4$ monolayer. 1$a$, 1$c$, and 1$e$ are \emph{Wyckoff} sites, and charges localize at \emph{Wyckoff} sites 1$a$ and 1$c$. The diamond solid line are the primitive cell of $\alpha_2$-MoSi$_2$N$_4$ monolayer. The charges distribution of band representation \={E}$_2$@1a (c) and \={E}$_2$@1e (d) of $\alpha_2$-MoSi$_2$N$_4$. The green hook face is the isosurface with value of 0.008. The black circles are OWCCs at 1$a$ and 1$e$.}
\label{alpha2-MoSi2N4}
\end{center}
\end{figure*}

\begin{table*}[htp]
\setlength{\tabcolsep}{3.5mm}
\begin{center}
\caption{All possible decompositions of the BR of $\alpha_1$-MoSi$_2$N$_4$ into linear combination of the EBRs in double space group  P$\bar{6}$m2 (No. 187). The first column gives the EBRs induced from different orbitals at Wyckoff positions 1a, 1c, and 1e ; the numbers below are the multiplicities of each EBR in the corresponding decomposition.}
\begin{tabular}{cccccccccc}
\hline\hline
No.&  \={E}$_1$@1c & \={E}$_2$@1c & \={E}$_3$@1c  & \={E}$_1$@1a & \={E}$_2$@1a & \={E}$_3$@1a  & \={E}$_1$@1e & \={E}$_2$@1e & \={E}$_3$@1e   \\\hline
\#01  &6 &7 &5 &1 &0 &0 &1 &1 &0  \\
\#02  &5 &6 &4 &2 &1 &1 &1 &1 &0  \\
\#03  &4 &5 &3 &3 &2 &2 &1 &1 &0  \\
\#04  &3 &4 &2 &4 &3 &3 &1 &1 &0  \\
\#05  &2 &3 &1 &5 &4 &4 &1 &1 &0  \\
\#06  &1 &2 &0 &6 &5 &5 &1 &1 &0  \\
\#07  &5 &6 &4 &1 &0 &0 &2 &2 &0  \\
\#08  &4 &5 &3 &2 &1 &1 &2 &2 &1  \\
\#09  &3 &4 &2 &3 &2 &2 &2 &2 &1  \\
\#10  &2 &3 &1 &4 &3 &3 &2 &2 &1  \\
\#11  &1 &2 &0 &5 &4 &4 &2 &2 &1  \\
\#12  &4 &5 &3 &1 &0 &0 &3 &3 &2  \\
\#13  &3 &4 &2 &2 &1 &1 &3 &3 &2  \\
\#14  &2 &3 &1 &3 &2 &2 &3 &3 &2  \\
\#15  &1 &2 &0 &4 &3 &3 &3 &3 &2  \\
\#16  &3 &4 &2 &1 &0 &0 &4 &4 &3  \\
\#17  &2 &3 &1 &2 &1 &1 &4 &4 &3  \\
\#18  &1 &2 &0 &3 &2 &2 &4 &4 &3  \\
\#19  &2 &3 &1 &1 &0 &0 &5 &5 &4  \\
\#20  &1 &2 &0 &2 &1 &1 &5 &5 &4  \\
\#21  &1 &2 &0 &1 &0 &0 &6 &6 &5  \\\hline\hline
\end{tabular}
\end{center}
\end{table*}

\begin{table*}[htp]
\setlength{\tabcolsep}{3.5mm}
\begin{center}
\caption{ All possible decompositions of the BR of $\alpha_2$-MoSi$_2$N$_4$ into linear combination of the EBRs in double space group  P$\bar{6}$m2 (No. 187). The first column gives the EBRs induced from different orbitals at Wyckoff positions 1a, 1c, and 1e; the numbers below are the multiplicities of each EBR in the corresponding decomposition. }
\begin{tabular}{cccccccccc}
\hline\hline
No.&  \={E}$_1$@1c & \={E}$_2$@1c & \={E}$_3$@1c  & \={E}$_1$@1a & \={E}$_2$@1a & \={E}$_3$@1a  & \={E}$_1$@1e & \={E}$_2$@1e & \={E}$_3$@1e   \\\hline
\#01  &5 &6 &5 &2 &1 &0 &1 &1 &0  \\
\#02  &4 &5 &4 &3 &2 &1 &1 &1 &0  \\
\#03  &3 &4 &3 &4 &3 &2 &1 &1 &0  \\
\#04  &2 &3 &2 &5 &4 &3 &1 &1 &0  \\
\#05  &1 &2 &1 &6 &5 &4 &1 &1 &0  \\
\#06  &0 &1 &0 &7 &6 &5 &1 &1 &0  \\
\#07  &4 &5 &4 &2 &1 &0 &2 &2 &0  \\
\#08  &3 &4 &3 &3 &2 &1 &2 &2 &1  \\
\#09  &2 &3 &2 &4 &3 &2 &2 &2 &1  \\
\#10  &1 &2 &1 &5 &4 &3 &2 &2 &1  \\
\#11  &0 &1 &0 &6 &5 &4 &2 &2 &1  \\
\#12  &3 &4 &3 &2 &1 &0 &3 &3 &2  \\
\#13  &2 &3 &2 &3 &2 &1 &3 &3 &2  \\
\#14  &1 &2 &1 &4 &3 &2 &3 &3 &2  \\
\#15  &0 &1 &0 &5 &4 &3 &3 &3 &2  \\
\#16  &2 &3 &2 &2 &1 &0 &4 &4 &3  \\
\#17  &1 &2 &1 &3 &2 &1 &4 &4 &3  \\
\#18  &0 &1 &0 &4 &3 &2 &4 &4 &3  \\
\#19  &1 &2 &1 &2 &1 &0 &5 &5 &4  \\
\#20  &0 &1 &0 &3 &2 &1 &5 &5 &4  \\
\#21  &0 &1 &0 &2 &1 &0 &6 &6 &5  \\\hline\hline
\end{tabular}
\end{center}
\end{table*}

\begin{table*}[htp]
\setlength{\tabcolsep}{4mm}
\begin{center}
\caption{Summary of OAI features of 17 34-VEC MA$_2$Z$_4$ materials, 2H-MoS$_2$, and $\alpha$-InSe .}
\begin{tabular}{clccl}
\hline\hline
No.& compounds name  &    phase type & OAI or not    & OWCC      \\\hline
01& CrSi$_2$N$_4$    &  $\alpha_1$   &      Y        & 1a,1e    \\
02& MoSi$_2$N$_4$    &  $\alpha_1$   &      Y        & 1a,1e    \\
03& MoSi$_2$N$_4$    &  $\alpha_2$   &      Y        & 1a,1e    \\
04& WSi$_2$N$_4$     &  $\alpha_1$   &      Y        & 1a,1e    \\
05& MoGe$_2$N$_4$    &  $\alpha_1$   &      Y        & 1a       \\
06& WGe$_2$N$_4$     &  $\alpha_1$   &      Y        & 1a       \\
07& CrSi$_2$P$_4$    &  $\alpha_2$   &      Y        & 1a,1e    \\
08& MoSi$_2$P$_4$    &  $\alpha_2$   &      Y        & 1a,1e    \\
09& WSi$_2$P$_4$     &  $\alpha_2$   &      Y        & 1a,1e    \\
10& CrGe$_2$P$_4$    &  $\alpha_2$   &      Y        & 1a       \\
11& MoGe$_2$P$_4$    &  $\alpha_2$   &      Y        & 1a       \\
12& WGe$_2$P$_4$     &  $\alpha_2$   &      Y        & 1a       \\
13& MoSi$_2$As$_4$   &  $\alpha_2$   &      Y        & 1a,1e    \\
14& WSi$_2$As$_4$    &  $\alpha_2$   &      Y        & 1a,1e    \\
15& MoGe$_2$As$_4$   &  $\alpha_2$   &      Y        & 1a       \\
16& WGe$_2$As$_4$    &  $\alpha_2$   &      Y        & 1a       \\
17& WSn$_2$N$_4$     &  $\alpha_1$   &      Y        & 1a       \\
18& MoS$_2$          &  2H           &      Y        & 1c,1e    \\
19& InSe             &  $\alpha$     &      Y        & 1a,1c      \\
\hline\hline
\end{tabular}
\end{center}
\end{table*}

\begin{figure*}[htp]
\begin{center}
\includegraphics[height=0.50\textwidth]{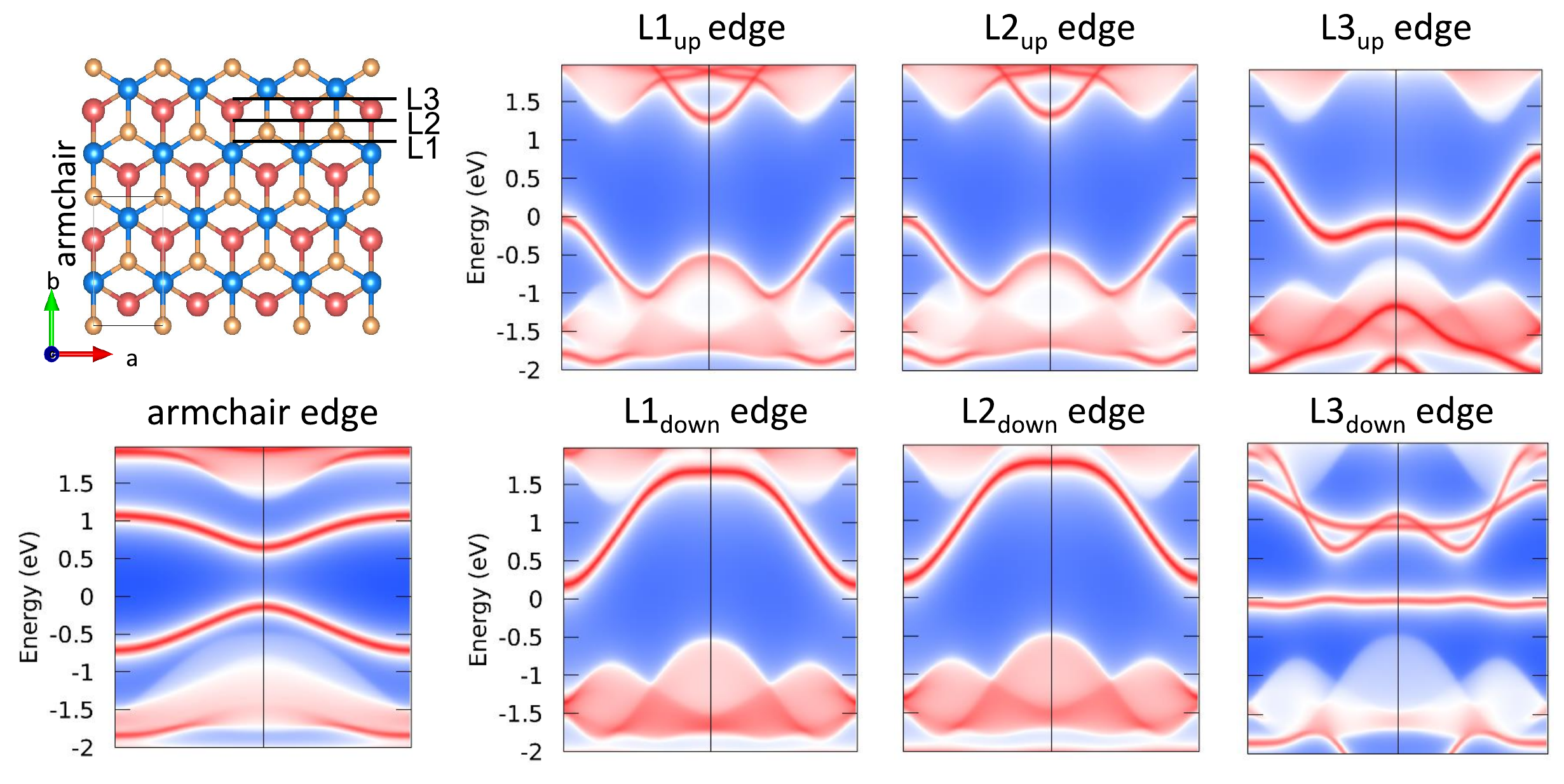}
\caption{(color online) The top left panel is the top view of MoSi$_2$N$_4$ monolayer. There are seven types of edges. Armchair edge is clear shown in figure. Another six edges, L$i_{up}$ and L$i_{down}$ ($i$= 1, 2, 3) edge, are obtained by three cutting lines, where subscript up and down are the edge above or below line L$_i$. The projected edge states for L$i_{up}$ and L$i_{down}$ ($i$= 1, 2, 3) and armchair edges are shown in another six panels.}
\label{phonons}
\end{center}
\end{figure*}

\begin{table*}[htp]
\setlength{\tabcolsep}{4mm}
\begin{center}
\caption{The structural information about nine C$_3$-symmetric nanodisks, where center type "2N" means two N atoms at the center of nanodisk and "2Si 2N" means two N atom and two Si atoms at the center of nanodisk; N$_{Mo}$, N$_{Si}$, and N$_{N}$ are the number of atoms of Mo, Si, and N in nanodisk; SR is the stoichiometric ratio in nanodisk, normalized down to N atoms.}
\begin{tabular}{ccccccccc}
\hline\hline
No.& C$_n$ & shape    & edge type & center type & N$_{Mo}$ & N$_{Si}$ & N$_{N}$ & SR  \\\hline
01& C$_3$ & hexagonal & armchair  & 2N          & 42       & 84       & 170     & 1.00:2.00:4.05  \\
02& C$_3$ & hexagonal & armchair  & 2Si 2N      & 42       & 86       & 170     & 1.00:2.05:4.05   \\
03& C$_3$ & hexagonal & armchair  & 1Mo         & 43       & 84       & 168     & 1.00:1.95:3.91   \\
04& C$_3$ & triangle  & armchair  & 2N          & 30       & 60       & 122     & 1.00:2.00:4.07   \\
05& C$_3$ & triangle  & armchair  & 2Si 2N      & 30       & 62       & 122     & 1.00:2.07:4.07   \\
06& C$_3$ & triangle  & armchair  & 1Mo         & 31       & 60       & 120     & 1.00:1.94:3.87   \\
07& C$_3$ & triangle  & L$2_{down}$ & 2N        & 15       & 42       & 98      & 1.00:2.80:6.53   \\
08& C$_3$ & triangle  & L$1_{down}$ & 2Si 2N    & 21       & 56       & 86      & 1.00:2.67:4.10  \\
09& C$_3$ & triangle  & L$3_{down}$ & 1Mo       & 28       & 30       & 72      & 1.00:1.07:2.57  \\
\hline\hline
\end{tabular}
\end{center}
\end{table*}

\begin{figure*}[htp]
\begin{center}
\includegraphics[height=0.70\textwidth]{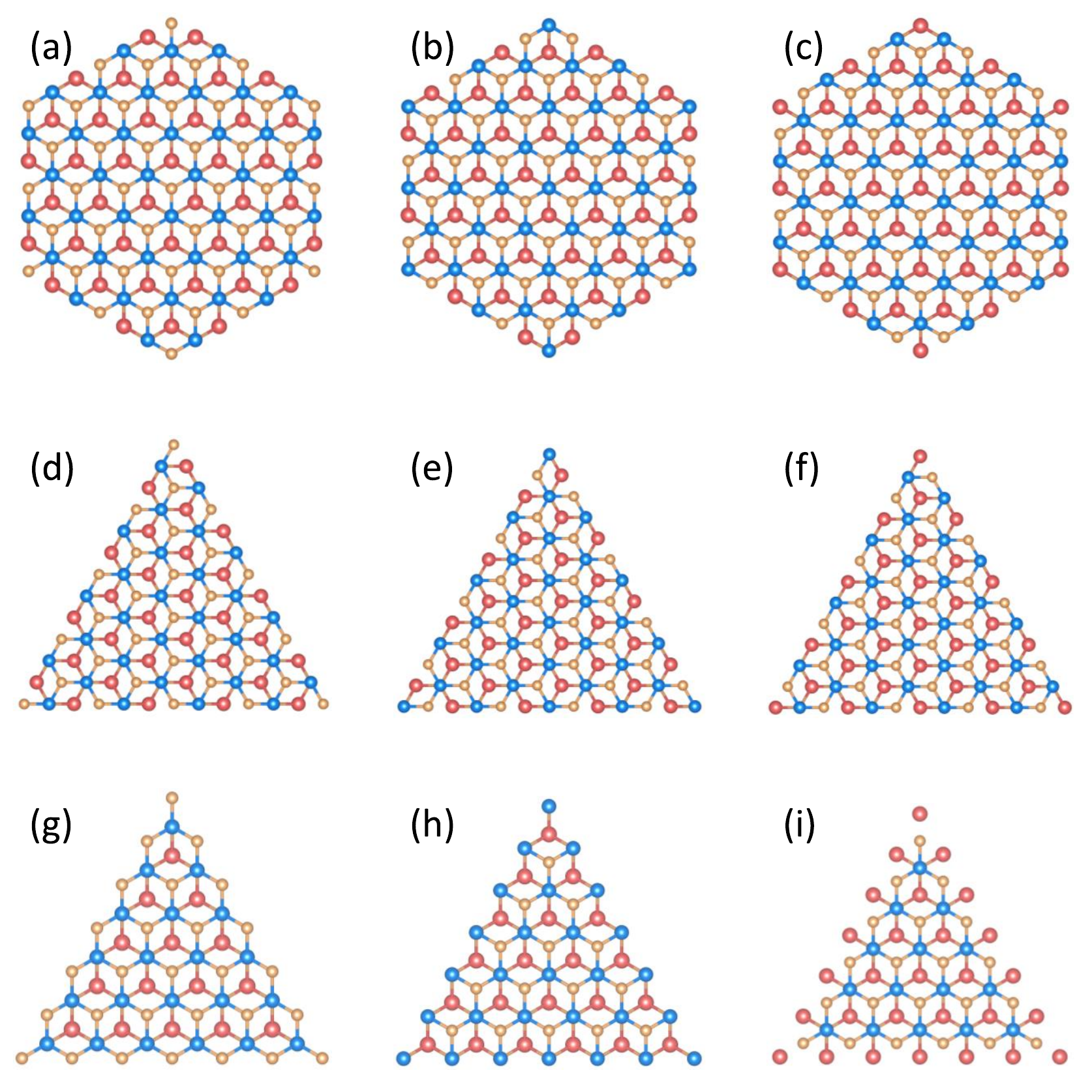}
\caption{(color online) Hexagonal and triangle C$_3$-symmetric nanodisk with armchair, and L$i_{down}$ ($i$ = 1, 2, 3) edge, corresponding to \textbf{Table A4}. }
\label{phonons}
\end{center}
\end{figure*}

\begin{figure*}[htp]
\begin{center}
\includegraphics[height=0.55\textwidth]{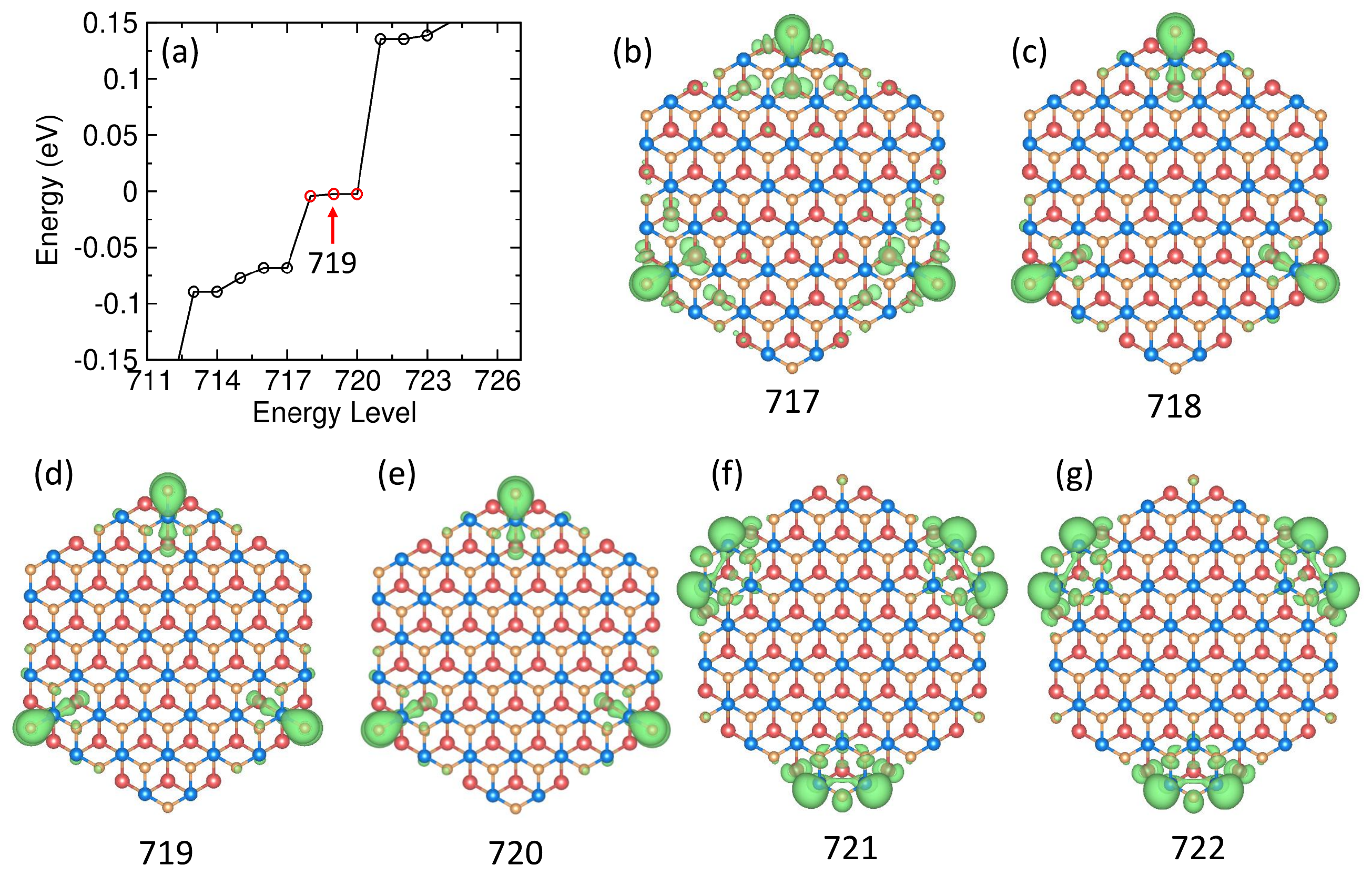}
\caption{(color online) (a) The energy spectrum of C$_3$-symmetric N-centered hexagonal nanodisk of MoSi$_2$N$_4$. Where red circles is corner states. (b-g) show the charge distribution of its corresponding energy level. }
\label{phonons}
\end{center}
\end{figure*}

\begin{figure*}[htp]
\begin{center}
\includegraphics[height=0.55\textwidth]{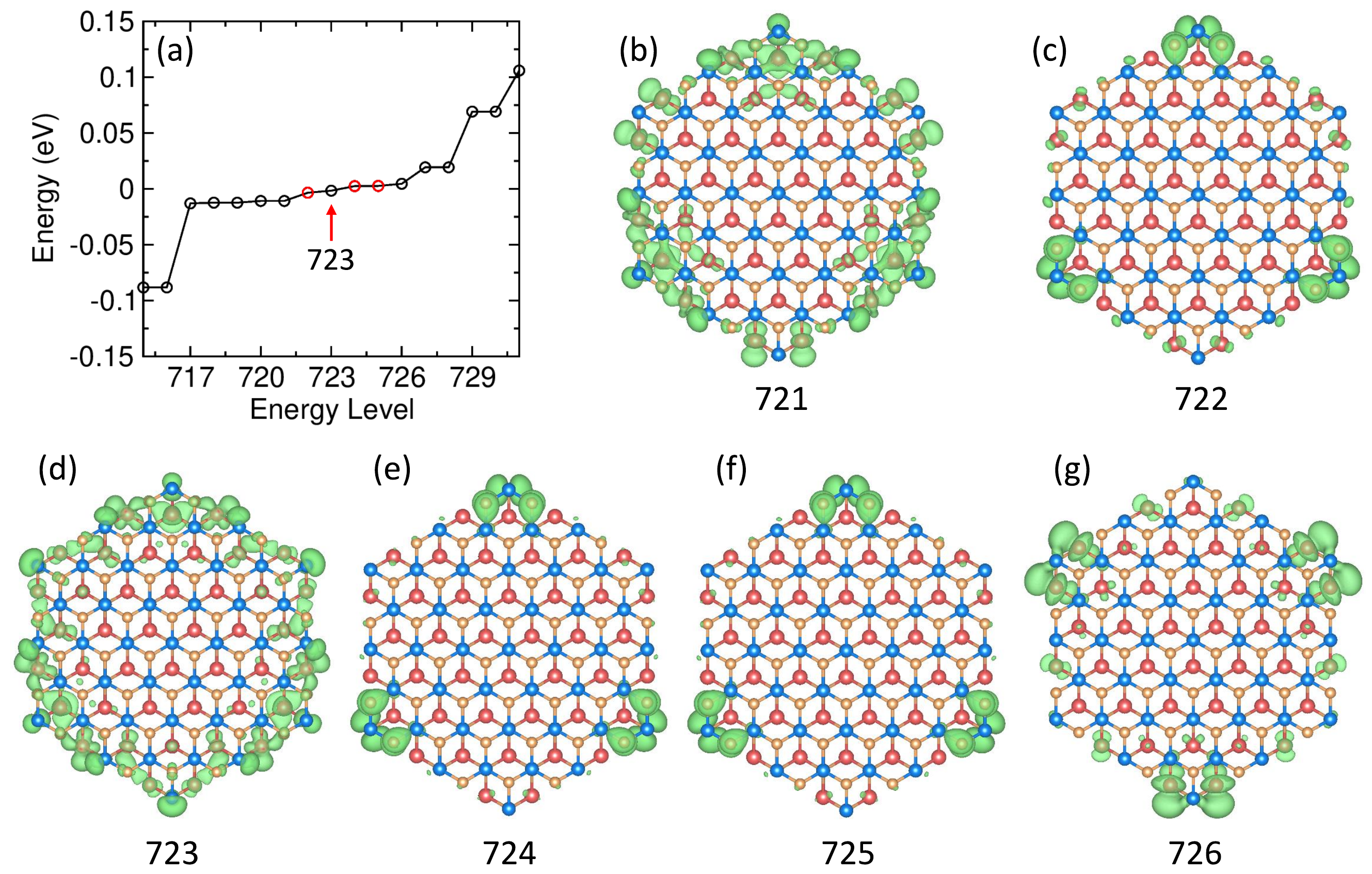}
\caption{(color online) (a) The energy spectrum of C$_3$-symmetric Si-centered hexagonal nanodisk of MoSi$_2$N$_4$. Where red circles is corner states. (b-g) show the charge distribution of its corresponding energy level.  }
\label{phonons}
\end{center}
\end{figure*}

\begin{figure*}[htp]
\begin{center}
\includegraphics[height=0.55\textwidth]{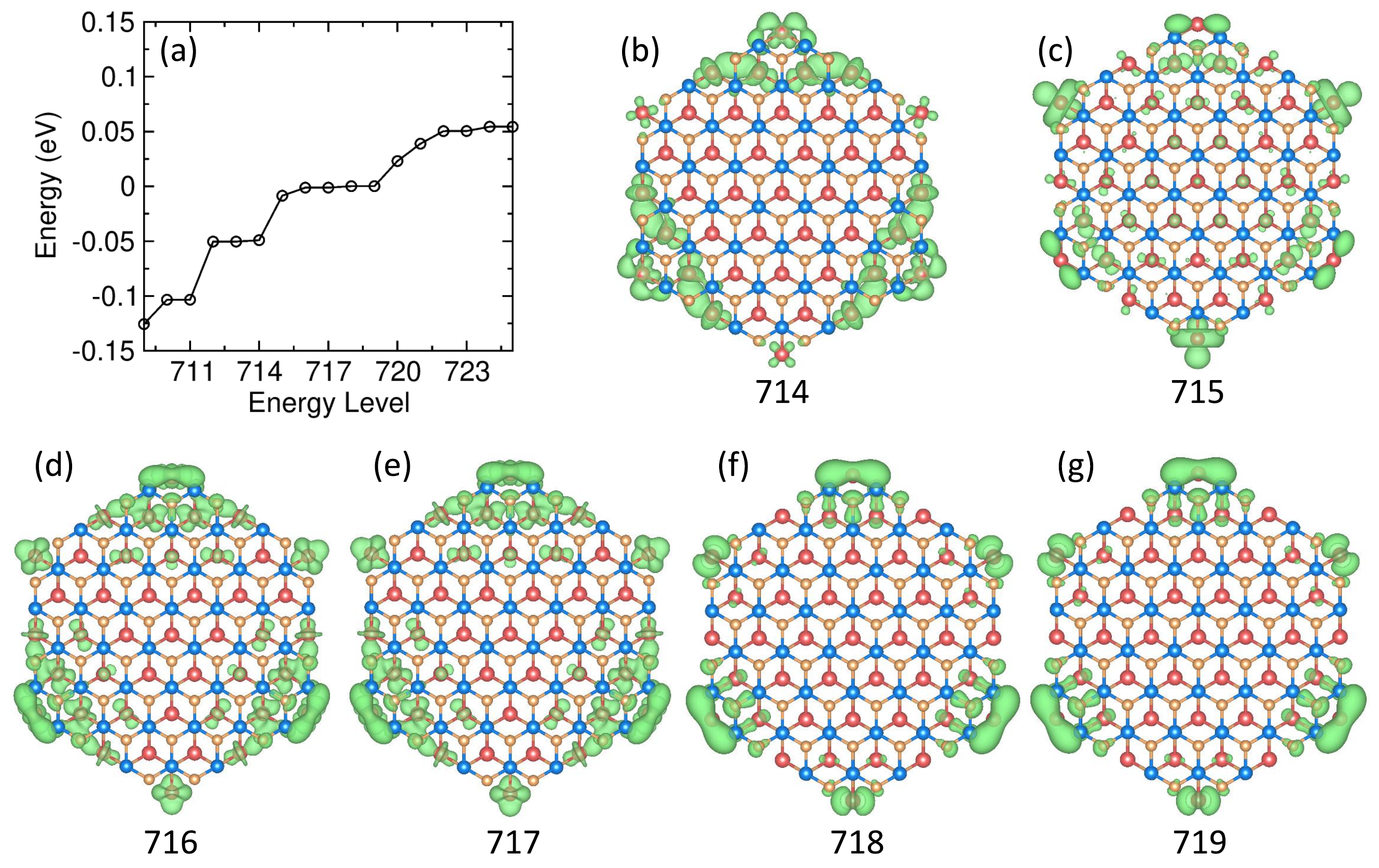}
\caption{(color online) (a) The energy spectrum of
C$_3$-symmetric Mo-centered hexagonal nanodisk of MoSi$_2$N$_4$. (b-g) show the charge distribution of
its corresponding energy level.}
\label{phonons}
\end{center}
\end{figure*}

\end{appendix}

\end{document}